\documentclass[prd,twocolumn,showpacs,nofootinbib,amsmath,amssymb,floatfix,superscriptaddress,showkeys]{revtex4}
\usepackage{multirow}
\usepackage{graphicx}
\usepackage{epstopdf}
\usepackage{dcolumn}
\usepackage{bm}
\usepackage{threeparttable}
\usepackage{subfigure}
\usepackage{color,txfonts}
\usepackage{ulem}
\definecolor{blue0}{rgb}{0,0,0.6}
\usepackage[colorlinks,linkcolor=blue0,anchorcolor=blue0,citecolor=blue0,urlcolor=blue0]{hyperref}

\newcommand{\beq}{\begin{equation}}
\newcommand{\eeq}{\end{equation}}
\newcommand{\beqa}{\begin{eqnarray}}
\newcommand{\eeqa}{\end{eqnarray}}

\begin{document}

\title{Search for the gamma-ray spectral lines with the DAMPE and the Fermi-LAT observations}
\author{Ji-Gui Cheng}
\affiliation{Guangxi Key Laboratory for Relativistic Astrophysics, School of Physical Science and Technology, Guangxi University, Nanning 530004, China}
\affiliation{School of Physics and Electronics, Hunan University of Science and Technology, Xiangtan 411201, China}
\author{Yun-Feng Liang}
\email{liangyf@gxu.edu.cn}
\affiliation{Guangxi Key Laboratory for Relativistic Astrophysics, School of Physical Science and Technology, Guangxi University, Nanning 530004, China}
\author{En-Wei Liang}
\email{lew@gxu.edu.cn}
\affiliation{Guangxi Key Laboratory for Relativistic Astrophysics, School of Physical Science and Technology, Guangxi University, Nanning 530004, China}

\date{\today}

\begin{abstract}
Weakly interacting massive particles, as a major candidate of dark matter (DM), may directly annihilate or decay into high-energy photons, producing monochromatic spectral lines in the gamma-ray band. These spectral lines, if detected, are smoking-gun signatures for the existence of new physics. Using the 5 years of DAMPE and 13 years of Fermi-LAT data, we search for line-like signals in the energy range of 3 GeV to 1 TeV from the Galactic halo. Different regions of interest are considered to accommodate different DM density profiles. We do not find any significant line structure, and the previously reported line-like feature at $\sim$133 GeV is also not detected in our analysis. Adopting a local DM density of $\rho_{\rm local}=0.4\,{\rm GeV\,cm^{-3}}$, we derive 95\% confidence level constraints on the velocity-averaged cross-section of $\langle{\sigma v}\rangle_{\gamma\gamma} \lesssim 4 \times 10^{-28}\,{\rm cm^{3}\,s^{-1}}$ and the decay lifetime of $\tau_{\gamma\nu} \gtrsim 5 \times 10^{29}\,{\rm s}$ at 100~GeV, achieving the strongest constraints to date for the line energies of 6-660 GeV. The improvement stems from the longer Fermi-LAT data set used and the inclusion of DAMPE data in the analysis. The simultaneous use of two independent data sets could also reduce the systematic uncertainty of the search.
\end{abstract}

\maketitle

\section{Introduction}
\label{sec:introduction}

Dark matter (DM; \cite{Jungman1996,Bertone2005}) dominates the matter component in the Universe. As suggested by the cosmic microwave background anisotropies, DM energy density is $\sim 6.4$ times that of baryon \citep{Planck2020}. Despite being abundant in the Universe, the nature of DM remains unknown. Many DM candidates have been proposed theoretically \citep{YoungBL2017}. As one group of the most widely studied DM candidates, Weakly interacting massive particles (WIMPs) can be found in many particle physics theories \citep{Feng2010}. They may produce monochromatic spectral lines in the gamma-ray band via annihilation or decay, e.g., $\chi \chi \to \gamma\gamma$ of supersymmetric particles \citep{Bergstrom1988} and $\chi \to \gamma\nu$ of gravitinos \citep{Ibarra2008}. Normal astrophysical processes are not expected to produce such sharp spectral lines. Therefore, the detection of line signals would be a smoking-gun signature for the existence of WIMPs.

Many research works have been carried out to search for spectral lines using gamma-ray observations from the EGRET, Fermi-LAT, DAMPE, HESS, and MAGIC \citep{Pullen2007,Abdo2010,Ackermann2012,Bringmann2012,Weniger2012,Geringer-Sameth2012,Huang2012,Tempel2012,Hektor2013,Ackermann2013,Albert2014,Ackermann2015,Liang2016,Anderson2016,hess16_line,hess18line,Li2019,HAWC:2019jvm,Shen2021,Alemanno2021,Liu2022,MAGIC:2022acl}. Among these works, several line candidates were temporarily reported, such as the $\sim 133$ GeV line signals detected from the Galactic center and galaxy clusters \citep{Bringmann2012,Weniger2012,Geringer-Sameth2012,Huang2012,Tempel2012,Ackermann2013,Hektor2013}, and the $\sim 43$ GeV tentative line signal found in the joint analysis of 16 nearby galaxy clusters \citep{Liang2016,Shen2021}. However, these line candidates are absent in the subsequent searches or do not have large enough significance \citep{Ackermann2015,Alemanno2021,Shen2021}. So far, no robust detection of a gamma-ray line has been claimed. Consequently, upper limits on the velocity-averaged cross-section $\langle{\sigma v}\rangle_{\gamma\gamma}$ and lower limits on the decay lifetime $\tau_{\gamma\nu}$ are placed for WIMPs. As reported in \cite{Ackermann2015}, constraints of $\langle{\sigma v}\rangle_{\gamma\gamma} \lesssim 10^{-26}\,{\rm cm^{3}\,s^{-1}}$ and $\tau_{\gamma\nu} \gtrsim 10^{28}\,{\rm s}$ at 100~GeV can be derived from the 5.8-year Fermi-LAT observation of the Galactic halo.

Similarly, the DAMPE collaboration has conducted a systematic search for line-like signals targeting the Galactic halo region with 5 years of accumulated data \citep{Alemanno2021}. DAMPE is very suitable for searching for sharp gamma-ray structures in GeV -- TeV range. It has an unprecedentedly-high energy resolution ($\sim 1.0\%$ at 100 GeV \cite{Chang2017}), which helps reduce the systematic uncertainty associated with background modeling. Ref.~\cite{Alemanno2021} has shown that, although having a much smaller effective area, the constraints on the DM parameters of producing gamma-ray lines based on DAMPE observations are comparable to that of \cite{Ackermann2015}.

Inspired by the previous studies, in this paper, we analyze the 5 years of DAMPE and 13 years of Fermi-LAT publicly available data from the Galactic halo to perform the line signal searches. The improvements of this work with respect to previous studies include: 1) we update the previous Fermi-LAT analysis (5.8 yr) with a longer dataset (13 yr); 2) we perform searches for line signals with the DAMPE and Fermi-LAT data jointly, which may reduce the uncertainties due to instrumental effects.

\section{Data reduction}
\label{sec:data}

The Fermi-LAT is a wide field-of-view imaging gamma-ray telescope launched in 2008 \citep{Atwood2009}. So far, over 14 years of observation data have been accumulated. In this work, we analyze the Fermi-LAT PASS 8 data ranging from 2008 August 4 to 2021 September 24 (MET: 239557417--654145676) to search for gamma-ray line signals. The latest (Ver. 2.2.0) {\tt Fermitools} software is used to process the data in the energy range 3-1000 GeV. We use the instrument response functions (IRFs) {\tt P8R3\_CLEAN\_V3} and corresponding LAT events ({\tt evclass=256} and {\tt evtype=3}) in our analysis. A maximum zenith angle of $100^{\circ}$ is set to eliminate the contamination from the bright Earth Limb. To ensure the data quality, the data quality cut {\tt (DATA\_QUAL>0)\&\&(LAT\_CONFIC==1)} is applied.

We adopt four different regions of interest (ROIs) with radii of $16^{\circ}$, $40^{\circ}$, $86^{\circ}$, and $150^{\circ}$ (denoted as R16, R40, R86, and R150 respectively) centering at the Sgr ${\rm A^{\ast}}$ (${\rm R.A.} = 266.415^{\circ}$, ${\rm Decl.} = -29.006^{\circ}$). The 4 ROIs are devised to optimize sensitivity for different DM density profiles \citep{Alemanno2021}. Since the Galactic plane is expected to be dominated by standard astrophysical processes, the region of ($|l|<\Delta l$ and $|b|<5^\circ$) is masked from the ROIs to obtain a better line search sensitivity, where the $\Delta l$ takes $5^{\circ}$, $9^{\circ}$, $0^{\circ}$ and $0^{\circ}$ for R16, R40, R86 and R150, respectively \cite{Alemanno2021}. We do not mask the photons from any detected point sources, since they only slightly affect the results (see Appendix A for details).

The DArk Matter Particle Explorer (DAMPE) is a space-based telescope launched in 2015 aiming to detect charged cosmic rays and gamma rays. It has great potential in searching for line signals owing to its unprecedentedly-high energy resolution \citep{Chang2017}. In this work, we use 5 years of DAMPE publicly-available data (from January 1, 2016 to January 1, 2021) to perform the line signal search. The {\tt DmpST} package \cite{Duan2019} is employed to analyze the data. For DAMPE data, the same ROIs are adopted as that used in the Fermi-LAT data reduction. Events belonging to both the Low Energy Trigger (LET) and the High Energy Trigger (HET) \citep{Duan2019} in the energy range of 3 GeV - 1 TeV are used. The events' incidence angles are restricted to $0.5 \le {\rm cos}(\theta) \le 1.0$ to ensure the data quality.

\section{Likelihood fitting}
\label{sec:data_fitting}

We perform an unbinned likelihood analysis in sliding energy windows to search for spectral lines \citep{Bringmann2012,Weniger2012,Ackermann2013,Albert2014,Ackermann2015,Liang2016}. The analysis procedure is briefly described below. We choose a series of line energies $E_{\rm line}$ with step lengths determined by the $68\%$ instrument energy resolution $\sigma_{E}$. The window width is defined as $[0.5 E_{\rm line}, 1.5 E_{\rm line}]$. We ignore the energy windows with event counts $n_{\rm evt}<30$ to guarantee sufficient statistics. Within each window, the maximum likelihood estimation is conducted by assuming a power-law background. The small width of the energy window warrants the power-law is a good approximation to the background spectrum (see Appendix B for the spectra of the 4 ROIs). Effects caused by the power-law approximation will be corrected by a systematic uncertainty term in the likelihood (see below).

For the null and signal models, the log-likelihood functions are respectively written as
    \begin{equation}
    \label{eq:lnLnull}
    {\rm ln}{\mathcal L}_{\rm null}(\theta_{\rm b}) = \sum_{i = 1}^{N} {\rm ln} (F_{\rm b}(E_{i};\theta_{\rm b}) \bar{\epsilon}(E_{i})) - \int F_{\rm b}(E;\theta_{\rm b}) \bar{\epsilon}(E) {\rm d}E
    \end{equation}
    and
    \begin{equation}
    \label{eq:lnLsig}
    \begin{split}
    {\rm ln}{\mathcal L}_{\rm sig}(N_{\rm s},E_{\rm line},\theta_{\rm b}) = \sum_{i = 1}^{N} {\rm ln} [(F_{\rm b}(E_{i};\theta_{\rm b}) \bar{\epsilon}(E_{i})) + F_{\rm s}(E_{i})\bar{\epsilon}(E_{\rm line})] \\
    - \int [F_{\rm b}(E;\theta_{\rm b}) \bar{\epsilon}(E) + F_{\rm s}(E) \bar{\epsilon}(E_{\rm line})] {\rm d}E.
    \end{split}
    \end{equation}
In the above equations, $E_{i}$ is the energy of each detected photon, $F_{\rm b}$ is the background flux, $F_{\rm s}$ is the line component expressed as $F_{\rm s} = N_{\rm s} \bar{D}(E;E_{\rm line})$ with $\bar{D}(E;E_{\rm line})$ the exposure-averaged instrument energy dispersion, and $\bar{\epsilon}$ is the average instrument exposure over the ROI. For details of our calculation of the energy dispersion $\bar{D}(E;E_{\rm line})$, please see Refs.~\cite{Liang2016} and \cite{Liu2022} for the Fermi-LAT and DAMPE, respectively. The $\theta_{\rm b}$ represents the nuisance parameters of the background. For a joint analysis of multiple data sets, the joint log-likelihood is the sum of individual log-likelihood values of each data set,
    \begin{equation}
    \begin{aligned}
    {\rm ln}{\mathcal L}_{\rm joint}&(N_{\rm s},E_{\rm line},\theta_{\rm b,1},\theta_{\rm b,2})= \\
    & {\rm ln}{\mathcal L}_{\rm F}(N_{\rm s},E_{\rm line},\theta_{\rm b,1})+{\rm ln}{\mathcal L}_{\rm D}(N_{\rm s},E_{\rm line},\theta_{\rm b,2}).
    \end{aligned}
    \end{equation}
    During the analysis, the fitting is implemented by the Python package {\tt iminuit}~\citep{iminuit}.

The local significance of a line signal can be approximated as the square root of the test statistic (TS) value, where ${\rm TS} \triangleq 2 ({\rm ln} {\mathcal L}_{\rm sig} - {\rm ln} {\mathcal L}_{\rm null})$. If no signal with ${\rm TS} > 25$ is found, the upper limit of $N_{\rm sig}$ can be derived by varying the best-fit log-likelihood value ${\rm ln}{\mathcal L}_{\rm sig}$ by 1.35.

It has been discussed that systematic uncertainties may induce a false line signal or mask a true one in the fitting \cite{Ackermann2013}. To account for such uncertainties, we adopt the same methodology as in \cite{Ackermann2015} to reform the likelihood equations. By replacing $N_{\rm s}$ with $N_{\rm sys} + N_{\rm sig}$, the systematic uncertainty term
\begin{equation}
{\rm ln}{\mathcal L}_{\rm sys}(n_{\rm sys}) = {\rm ln} \left(\frac{1}{\sqrt{2\pi} \sigma_{\rm sys}} e^{-n_{\rm sys}^{2} / 2 \sigma_{\rm sys}^{2}}\right)
\label{eq:gauss}
\end{equation}
is added into Eq.~(\ref{eq:lnLnull}) and Eq.~(\ref{eq:lnLsig}). In the above equation, the fractional signal is defined as $f \equiv n_{\rm sig} / b_{\rm eff}$ so that $\delta f_{\rm sys}$ can be obtained in the fitting of control regions ($\delta f_{\rm sys} \lesssim 1.5\%$ or $2.0\%$ for the Fermi-LAT and the DAMPE respectively, referring to \cite{Ackermann2015} and \cite{Alemanno2021}), and the effective background $b_{\rm eff}$ is the number of background photon counts under the signal peak \citep{Ackermann2015}. More details on the effective background and the fitting of control regions are given in Appendix C and D, respectively.

\begin{figure}
\centering
\includegraphics[width=0.98\columnwidth]{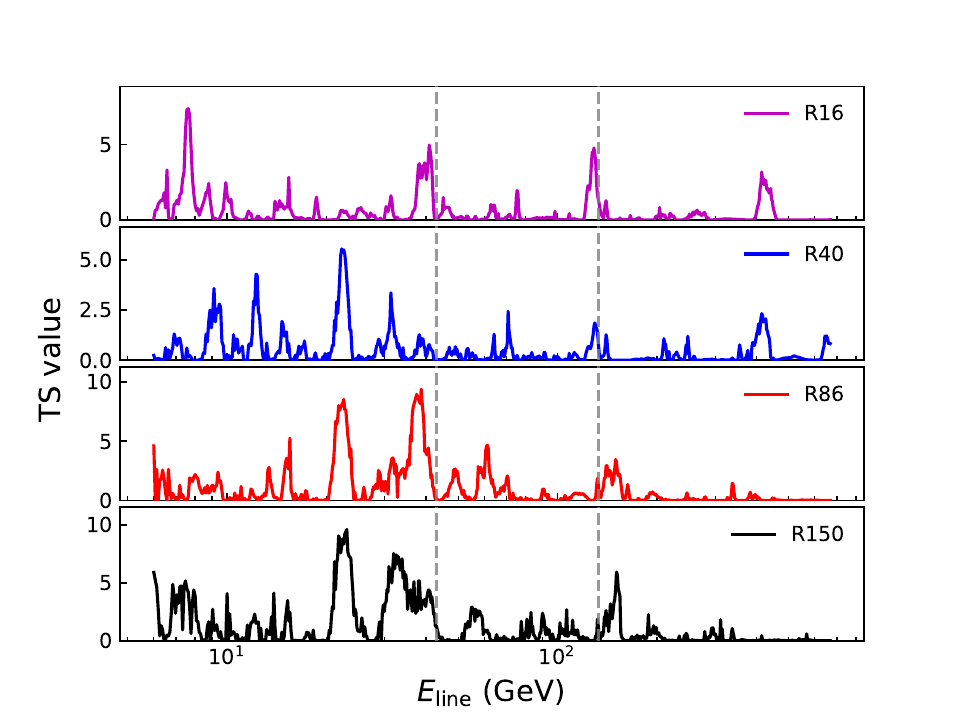}
\caption{TS values of putative gamma-ray lines as a function of line energies obtained in the joint likelihood analysis. From top to bottom, the panels are for 16$^\circ$, 40$^\circ$, 86$^\circ$ and 150$^\circ$ ROIs centered on the Galactic center, respectively. No line candidate is found with ${\rm TS_{joint} > 25}$. The two vertical dashed lines represent the tentative line signals at $\sim43$ GeV \citep{Liang2016} and $\sim133$ GeV \citep{Bringmann2012,Weniger2012}, respectively.}
\label{fig:ts}
\end{figure}

\section{Predicted flux of DM spectral line}
\label{sec:model}

Following previous studies of line signal searches \citep{Ackermann2015,Alemanno2021}, we consider three commonly used DM density profiles, which are (1) the Navarro-Frenk-White (NFW) profile \citep{Navarro1996}
\begin{equation}
\rho_{\rm NFW}(r) =\frac{\rho_{\rm s}}{(r / r_{\rm s})(1 + r / r_{\rm s})^{2}}
\end{equation}
with $r_{\rm s} = 20\,{\rm kpc}$; (2) the Einasto profile \citep{Einasto1965,Navarro2010}
\begin{equation}
\rho_{\rm Ein}(r) = \rho_{\rm s} {\rm exp}\{-(2 / \alpha) [(r / r_{\rm s})^{\alpha} - 1]\}
\end{equation}
with $r_{\rm s} = 20\,{\rm kpc}$ and $\alpha = 0.17$; (3) the isothermal profile \citep{Bahcall1980}
\begin{equation}
\rho_{\rm iso}(r) = \frac{\rho_{\rm s}}{1 + (r / r_{\rm s})^{2}}
\end{equation}
with $r_{\rm s} = 5\,{\rm kpc}$. The local DM density $\rho_{\rm local}$ is set to $\rho_{\rm DM}(R_{\rm 0}) = 0.4\,{\rm GeV\,cm^{-3}}$ with $R_{\rm 0} = 8.5\,{\rm kpc}$. Note that this value of $\rho_{\rm local}$ is intermediate among the ones reported in the literature \cite{deSalas:2020hbh}. A higher or lower value would strengthen or weaken the constraints on the DM parameters.

For each DM density profile, the J-factor and D-factor is calculated through
$J_{\rm DM} = \int_{\rm ROI} {\rm d}\Omega \int {\rm d}l\,\rho_{\rm DM}^{2}$
and
$D_{\rm DM} = \int_{\rm ROI} {\rm d}\Omega \int {\rm d}l\,\rho_{\rm DM}$,
respectively. The ROIs and the J/D-factors are paired into (R16, $J_{\rm Ein}$), (R40, $J_{\rm NFW}$), (R86, $J_{\rm iso}$) and (R150, $D_{\rm NFW}$) with corresponding values referring to the Table 1 of Ref.~\cite{Alemanno2021}.

The expected flux from DM annihilation $\chi \chi \to \gamma \gamma$ is expressed as
\begin{equation}
\label{eq:Sline_ann}
\begin{aligned}
S_{\rm line}(E) &= \frac{1}{4 \pi} \frac{\langle{\sigma v}\rangle_{\gamma\gamma}}{2 m_{\rm \chi}^{2}} 2 \delta(E - E_{\rm line}) \times J_{\rm DM}\\
&=N_{\rm sig} \delta(E - E_{\rm line}),
\end{aligned}
\end{equation}
in which $\langle{\sigma v}\rangle_{\gamma\gamma}$ is the velocity-averaged annihilation cross section, $m_{\rm \chi}$ is the DM mass, and the energy of line photons is $E_{\rm line} = m_{\rm \chi}$. The expected flux from DM decay $\chi \to \gamma \nu$ is given by
\begin{equation}
\label{eq:Sline_dec}
\begin{aligned}
S_{\rm line}(E) & = \frac{1}{4 \pi} \frac{1}{m_{\rm \chi} \tau_{\gamma \nu}} \delta(E - E_{\rm line}) \times D_{\rm DM}\\
&=N_{\rm sig} \delta(E - E_{\rm line}),
\end{aligned}
\end{equation}
where $\tau_{\gamma \nu}$ is the decay lifetime, and the energy of line signal is $E_{\rm line} = m_{\rm \chi} / 2$.

\begin{figure*}
\centering
\includegraphics[width=0.43\textwidth]{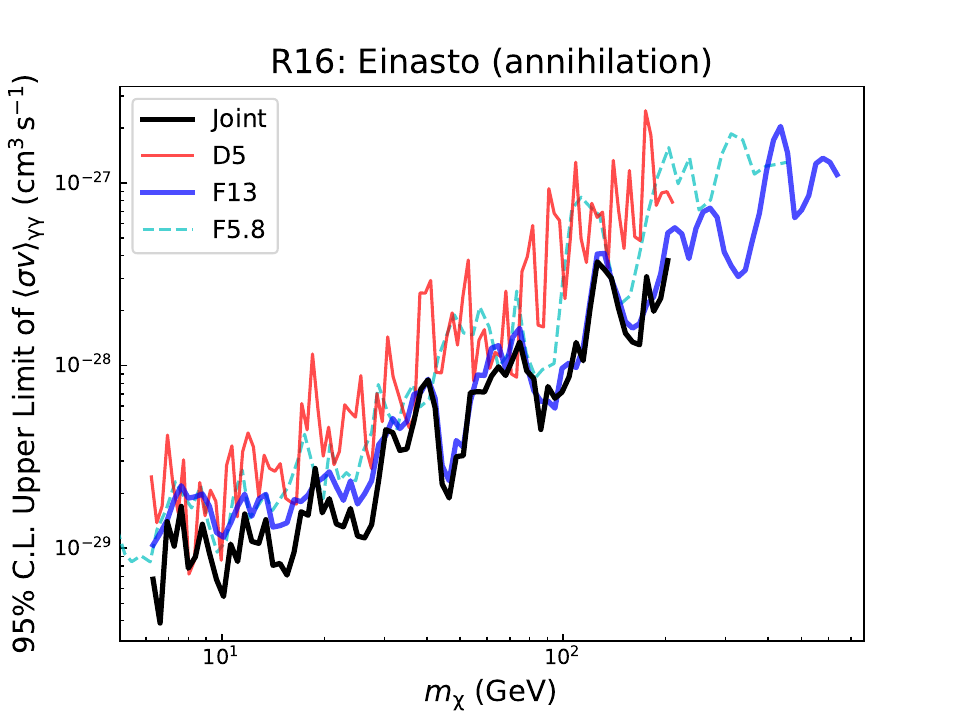}
\includegraphics[width=0.43\textwidth]{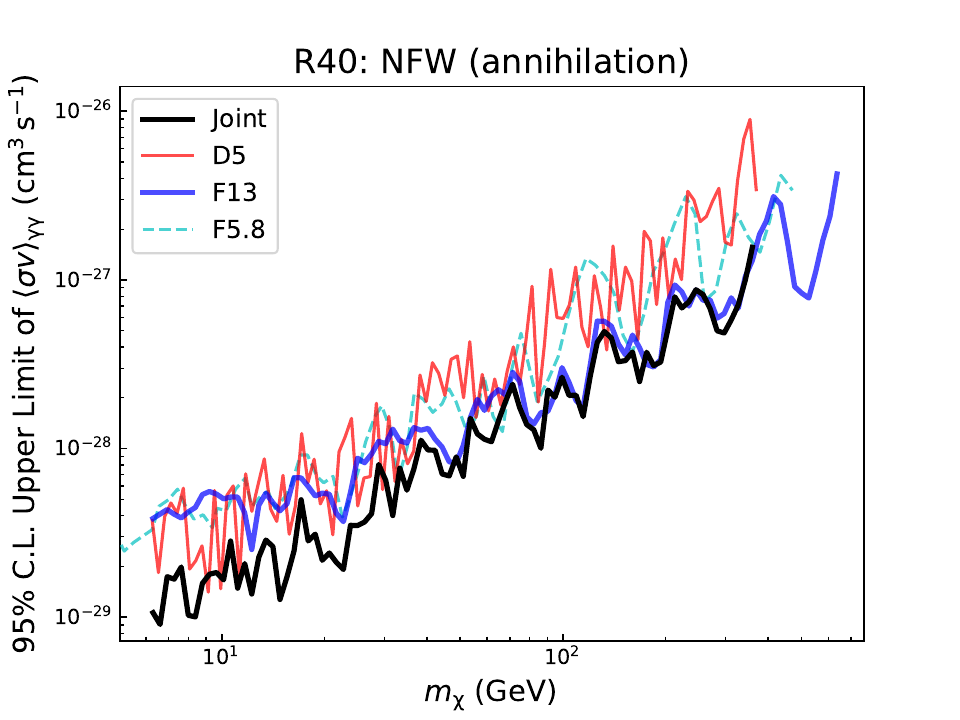} \\
\includegraphics[width=0.43\textwidth]{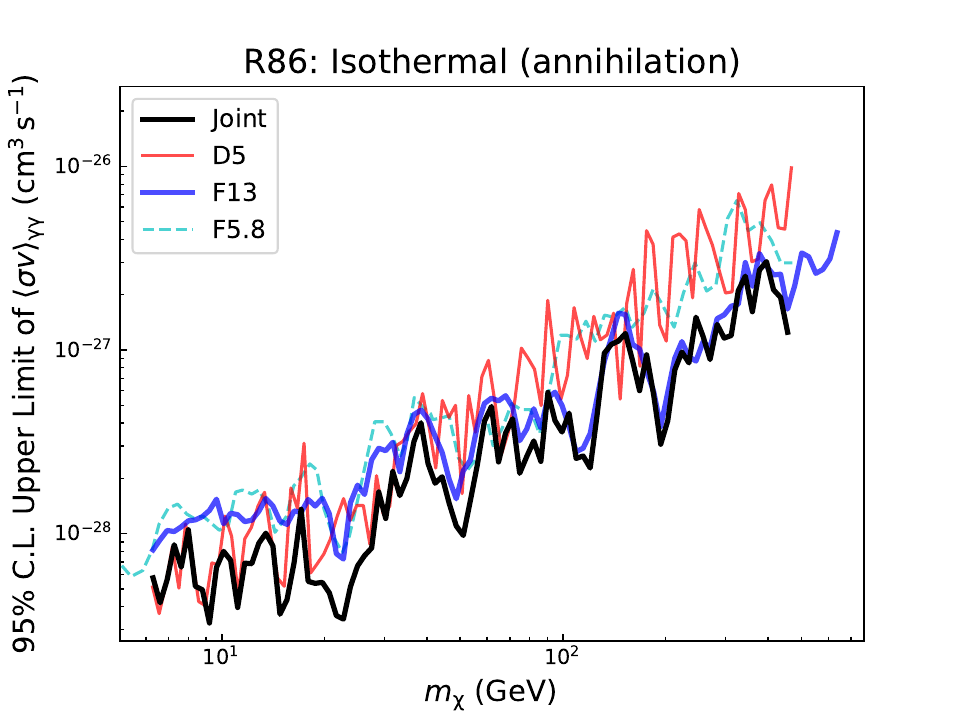}
\includegraphics[width=0.43\textwidth]{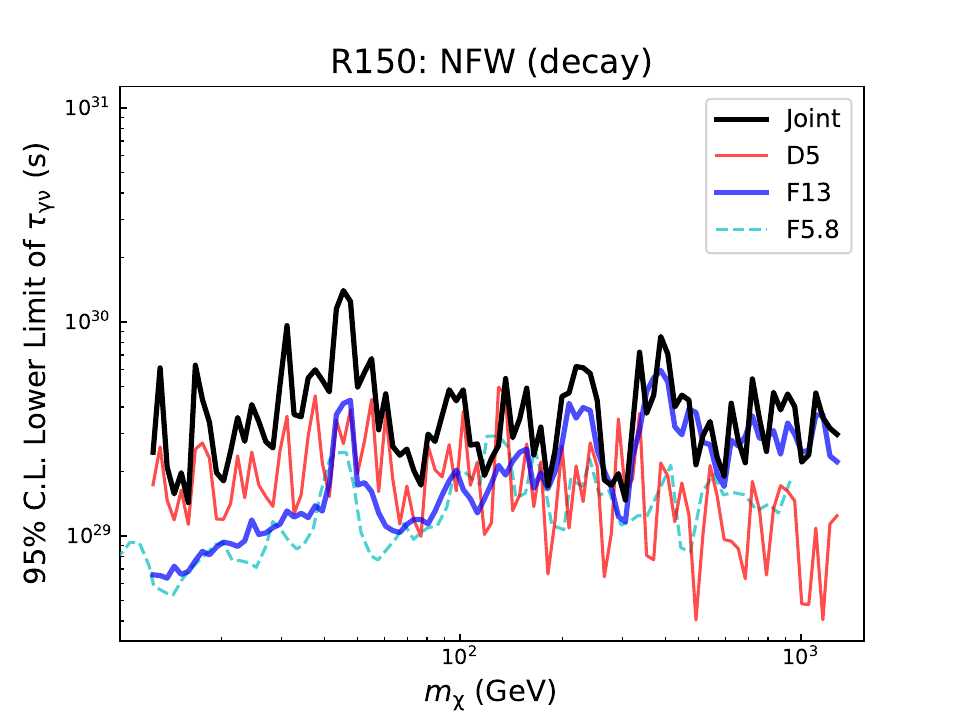} \\
\caption{The 95\% confidence level constraints on $\langle{\sigma v}\rangle_{\gamma\gamma}$ and $\tau_{\gamma \nu}$ for different ROIs and DM density profiles. The D5 and F13 are the constraints based on 5 years of DAMPE data and 13 years of Fermi-LAT data, respectively. The joint constraints are derived by combining the likelihood of D5 and F13 in the unbinned likelihood analysis. The limits reported in \cite{Ackermann2015} based on 5.8 years of Fermi-LAT data are also shown for comparison (denoted as F5.8).}
\label{fig:limits}
\end{figure*}

\section{Results and discussion}
\label{sec:results}

Assuming DM annihilates or decays through $\chi \chi \to \gamma \gamma$ and $\chi \to \gamma \nu$ channels, respectively, we perform line searches with the DAMPE and the Fermi-LAT data. The search results of our joint analysis of DAMPE and Fermi data are shown in Fig.~\ref{fig:ts}, where we show the TS values of potential line signals as a function of the line energy for the four selected ROIs. In total, there are 721 energy windows according to the DAMPE's energy resolution in the energy range from 3 to 1000 GeV. In these windows, we find that no line signals have ${\rm TS > 25}$ (corresponding to a local significance of $\gtrsim 5 \sigma$) for the DM mass range from $m_\chi \sim 6\,{\rm GeV}$ to 660 GeV\footnote{The range is a little narrower than that of our entire data sets (i.e., 3-1000 GeV) to allow for the spectral sidebands.}. For R86 and R150, the highest TS values are $\lesssim 10$, while for R16 and R40 the highest TS is merely $\sim 6$.
The previously reported tentative line signal at $\sim$133~GeV is not favored by the current joint analysis.

Since no line signal is found, we derive upper limits on $N_{\rm sig}$, and then convert them into the upper limits on $\langle{\sigma v}\rangle_{\rm \gamma\gamma}$ or lower limits on $\tau_{\rm \gamma \nu}$ via Eq.~(\ref{eq:Sline_ann}) and Eq.~(\ref{eq:Sline_dec}), respectively. As is shown in Fig.~\ref{fig:limits}, the 5 years of DAMPE data lead to constraints of $\langle{\sigma v}\rangle_{\gamma\gamma} \lesssim 9.0 \times 10^{-26}\,{\rm cm^{3}\,s^{-1}}$ and $\tau_{\gamma \nu} \gtrsim 3.0 \times 10^{28}\,{\rm s}$ at $m_\chi=100\,\rm GeV$ (denoted as D5), consistent with the previous results by the DAMPE Collaboration \cite{Alemanno2021}. Constraints derived from the 13 years of Fermi-LAT data (denoted as F13), on the other hand, are generally stronger than the D5 results, $\langle{\sigma v}\rangle_{\gamma\gamma,\rm 100GeV} \lesssim 5.0 \times 10^{-27}\,{\rm cm^{3}\,s^{-1}}$ and $\tau_{\gamma \nu,\rm 100GeV} \gtrsim 5.0 \times 10^{30}\,{\rm s}$.
We note that the constraints from the joint analysis of the Fermi-LAT and DAMPE data (black line) are influenced by both the Fermi-LAT and DAMPE data, and are stronger than both single constraints. Also, the joint constraints are more stringent than those from the 5.8 years of Fermi-LAT data (\cite{Ackermann2015}; denoted as F5.8) and the 5 years of DAMPE data \citep{Alemanno2021}. Our joint analysis, therefore, places the currently strongest constraints on gamma-ray lines from DM at $6-660$ GeV energies to date.

\begin{figure}
\includegraphics[width=0.9\columnwidth]{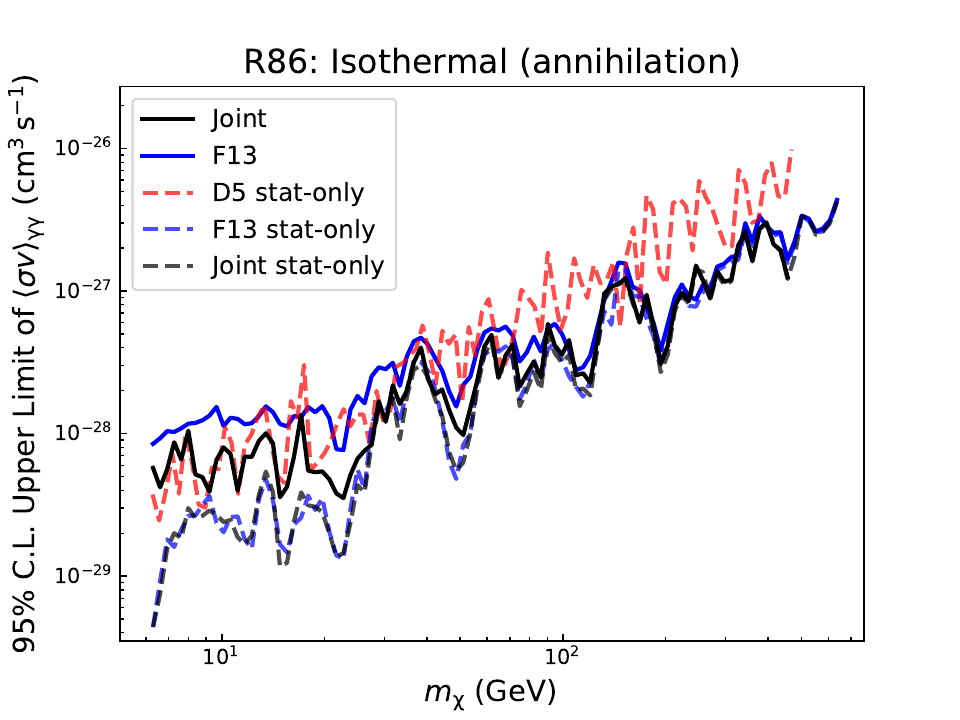}
\caption{Constraints on $\langle{\sigma v}\rangle_{\gamma\gamma}$ from the analysis {\it ignoring systematic uncertainties} and comparisons to our nominal results which take systematic effects into account. Only the results of the R86 ROI are shown here as a demonstration, see Appendix G for other ROIs.}
\label{fig:stat}
\end{figure}

The above results have accounted for systematic uncertainties by including an additional Gaussian term (Eq.~\ref{eq:gauss}) in the likelihood. Fig.~\ref{fig:stat} presents the results that ignore the systematic uncertainties, which can help us to understand the contributions of the two data sets in the joint analysis.
It can be seen that the inclusion of systematic uncertainty prevents the Fermi-LAT results from improving with the data accumulation at the low-energy end. This is reasonable, because when the photon statistics is high enough, the detection sensitivity is dominated by the systematic uncertainty and will not change with the accumulation of data (see Appendix E for a detailed discussion). In contrast, although the DAMPE data are statistically subdominant, including DAMPE data can effectively reduce the sensitivity loss owing to the consideration of systematic uncertainty because of its higher energy resolution.

To further validate the improved results achieved through higher energy resolution, we conduct tests using the EDISP3 data of Fermi-LAT (a subgroup of data that has the best energy reconstruction quality, see Appendix F for details) to derive the constraints. We find that in the low-energy range, better results are obtained from the EDISP3 analysis, indicating the higher energy resolution (which reduces $b_{\rm eff}$ and $\delta f_{\rm sys}$ in Eq.~(\ref{eq:gauss})) does help to improve the analysis results. In the high-energy range the EDISP3 constraints are slightly weaker than our nominal results because the EDISP3 data lose 3/4 of the statistics.
Note that this paper does not separately consider the four EDISP event types of Fermi-LAT data to perform a summed likelihood analysis. If so, the high energy resolution of EDISP3 can be utilized without loss of statistics, and the results could be further improved. Previous analyses pointed out that it could improve $\sim15\%$ for the Fermi-LAT only analysis \cite{Ackermann2013,Ackermann2015}.

This work combines the DAMPE and Fermi-LAT data to perform spectral line searches, which to our knowledge is the first work that combines data from different gamma-ray detectors to search for DM signals.
Although the 130 GeV tentative signal is not supported by subsequent analyses \cite{Finkbeiner:2012ez,Ackermann2015,Alemanno2021}, the approach we show here is not limited to the searches of spectral lines.
Our work reveals the great potential of combining different instruments in improving the sensitivity of DM indirect search, demonstrating a new means to enhance the sensitivity.  In the future, with more and more gamma-ray detectors (such as HERD \cite{Huang:2015fca} and VLAST \cite{2022AcASn..63...27F}) starting observations, and with the increasing accumulation of Fermi-LAT and DAMPE data, such a joint analysis will become even more important.

\section{Summary}
\label{sec:sum}

In this work, we jointly analyze the Fermi-LAT and DAMPE data to search for line-like signals from the DM annihilation or decay in the Galactic halo. Using the sliding energy window method and unbinned likelihood analysis, we find no evidence for a gamma-ray line having a local significance of $> 5\sigma$. Assuming DM particles annihilate through $\chi \chi \to \gamma \gamma$ channel or decay through $\chi \to \gamma \nu$ channel and adopting a local DM density of $\rho_{\rm local}=0.4\,{\rm GeV\,cm^{-3}}$, we derive the upper limits on $\langle{\sigma v}\rangle_{\gamma\gamma}$ and the lower limits on $\tau_{\gamma \nu}$ for different ROIs and DM density profiles. We improve the previous constraints on these parameters by a factor of $\sim \sqrt{2}$.

The improvement is due in part to the use of larger Fermi-LAT dataset and in part to the combined analysis of data from two independent instruments.
The simultaneous use of the Fermi-LAT and DAMPE data makes the obtained constraints stronger than the ones based on Fermi-LAT or DAMPE data alone. Further, since two independent instruments are employed, the systematic effects from individual instruments could also be reduced.

Although line signals are still absent and the constraints are getting more stringent, WIMP is still one of the most promising DM models. Recently, two experiments at Fermilab, E989 and CDF II, reported anomalies for muon anomalous magnetic moment ($g-2$) \citep{Muong-2:2021ojo} and $W$-boson mass \citep{CDF2022}, deviating from the prediction of the Standard Model by about 4.2$\sigma$ and 7$\sigma$, respectively. Such results may indicate the existence of new physics and can be well explained by introducing WIMPs \citep{Fan:2022dck,Yuan:2022cpw,Zhu:2022tpr}, which motivates us to keep searching for the annihilation or decay signals from WIMPs (e.g., gamma-ray lines).

\begin{acknowledgments}
We thank Qiang Yuan, Yizhong Fan and Zhaoqiang Shen for the helpful suggestions and discussions. We acknowledge data and software provided by the Fermi Science Support Center. We acknowledge data resources from DArk Matter Particle Explorer (DAMPE) satellite mission supported by Strategic Priority Program on Space Science, and data service provided by National Space Science Data Center of China. This work is supported by the National Natural Science Foundation of China (No. 12133003), the Guangxi Science Foundation (No. 2019AC20334).
\end{acknowledgments}

\widetext
\newpage
\appendix

\section*{Supplemental Material}

\section{Effect of point sources on the analysis results}
\label{sec:psmask}
For DM searches, the known astrophysical gamma-ray sources constitute one of the backgrounds of the signal search. They could either overwhelm a real signal or produce a pseudo signal. However, for the searches of spectral lines, it is common practice not to model/remove any known gamma-ray point sources (Ref.~\cite{Ackermann2013} is one work that masked bright point sources in their analysis). This is because, considering the characteristic shape of the spectral line signal, the emission from any astrophysical point source is not expected to produce such a sharp line-like structure, but is continuous in energy. So in a small energy window the gamma-ray background that mixes all the diffuse and point sources can be approximated as a power-law spectrum, and we do not need to model each single source (i.e., the sliding energy window method). The advantage of this method is that it simplifies the analysis process, avoids too many background model parameters, and reduces the bias caused by the inaccurate assumption of background models.
Systematic effects such as false positive or negative signals (concave or bumps in the power-law background) caused by the power-law approximation of the background can be taken into account by adding a systematic uncertainty term into the likelihood fit.

\begin{figure*}[h]
\centering
\includegraphics[scale=0.4]{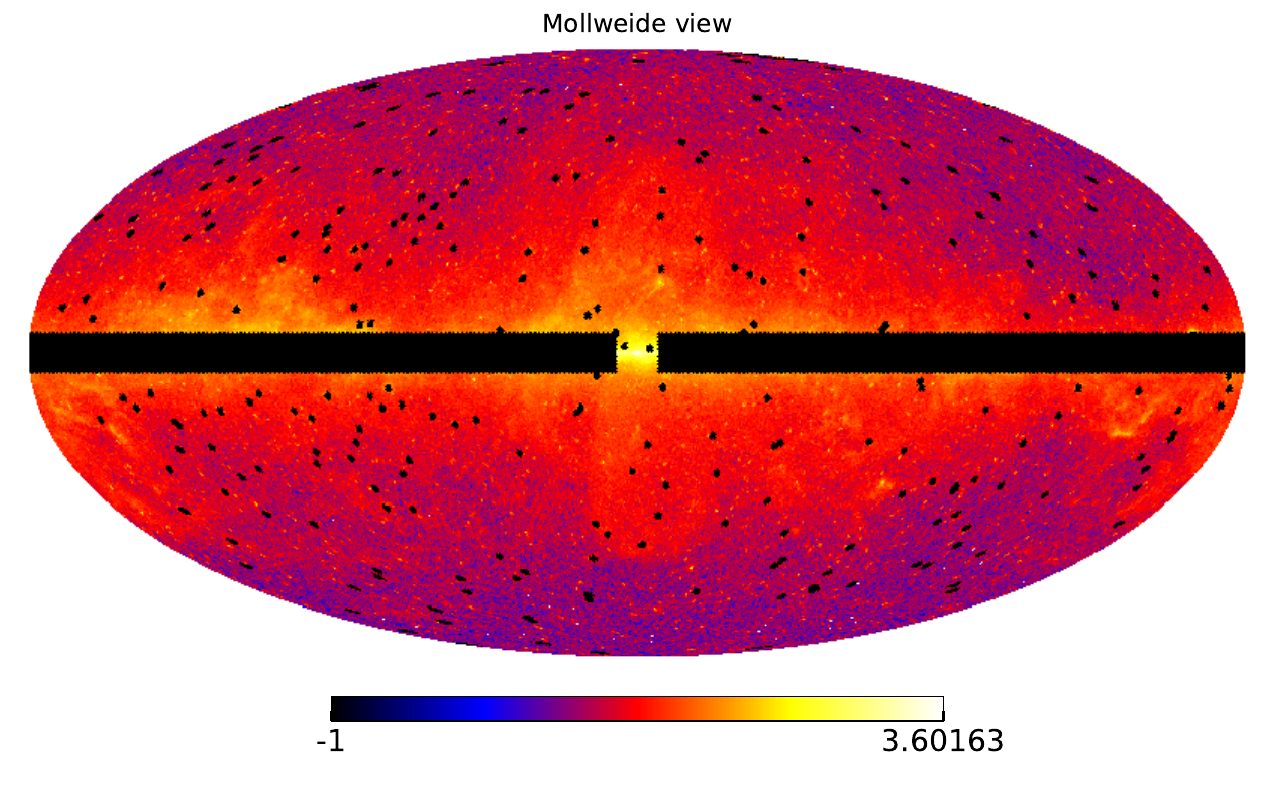}
\includegraphics[scale=0.4]{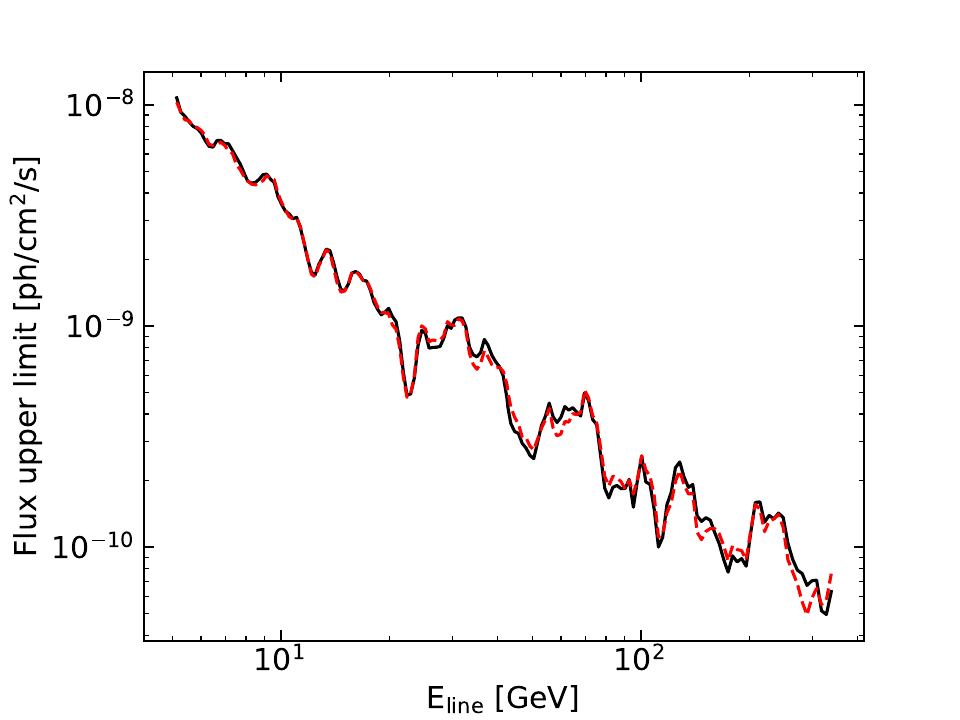}\\
\caption{Left: Fermi-LAT $>3\,{\rm GeV}$ all-sky counts map. The Galactic plane region and the 300 brightest point sources are masked. Right: A comparison of the results of R40 ROI based on the data with/without point source masking.}
\label{fig:mask}
\end{figure*}

Despite of this, here we also test whether masking point sources affects the results of our analysis. We perform this test only for the Fermi-LAT data and the NFW profile (R40 ROI). We sort the sources of the Fermi-LAT 4FGL-DR3 catalog \cite{Fermi-LAT:2019yla,Fermi-LAT:2022byn} by their significance and then mask out the 300 most bright sources from the all-sky data. Considering that the point spread function (PSF) of Fermi-LAT above 3 GeV is less than 1 degree \footnote{\url{https://www.slac.stanford.edu/exp/glast/groups/canda/lat_Performance.htm}}, we remove the photons within the 1$^\circ$ region around these sources from the data (see Fig.~\ref{fig:mask}). These regions have also been removed when computing the average exposure. Masking out these point source regions will reduce the J-factor by about 2.5\%, and the total amount of data in the R40 region will decrease by about 4.4\%. Using these data we perform exactly the same analysis as in the main text and get the flux upper limits as shown in the Fig.~\ref{fig:mask}. One can see that whether or not the point sources are masked will only have a very small effect on the results. In \cite{Ackermann2013}, they have also shown that the point source contamination results in a uncertainty of $|\delta f<0.05|$ and is sub-dominant compared to other systematic uncertainties.

\section{The spectra of the 4 ROIs}
\label{ap:4sp}
\begin{figure*}[h]
\centering
\includegraphics[scale=0.4]{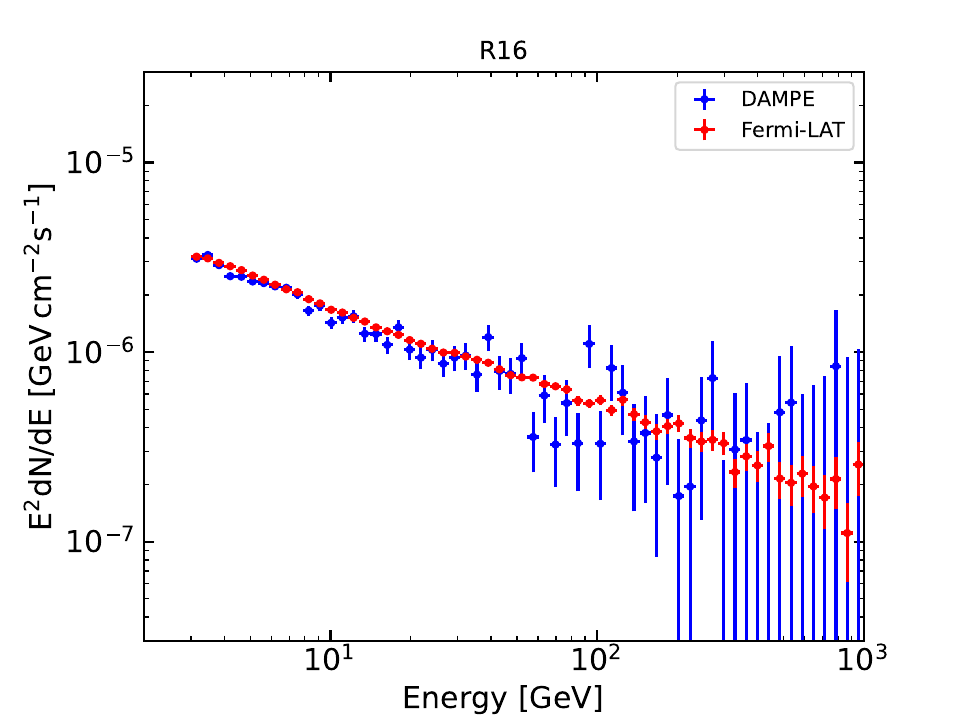}
\includegraphics[scale=0.4]{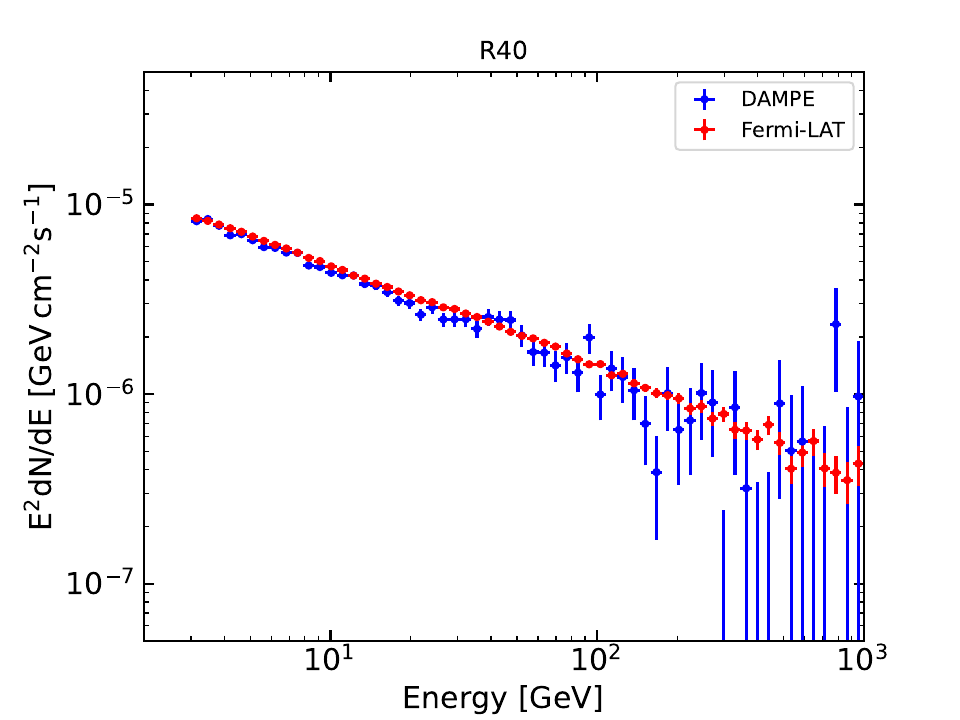}\\
\includegraphics[scale=0.4]{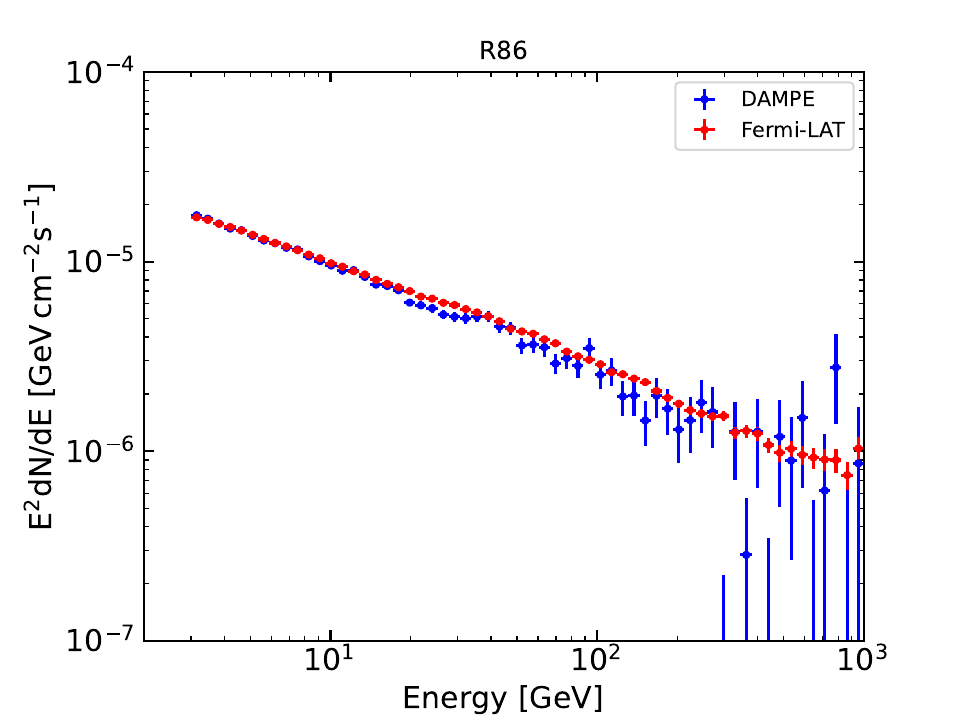}
\includegraphics[scale=0.4]{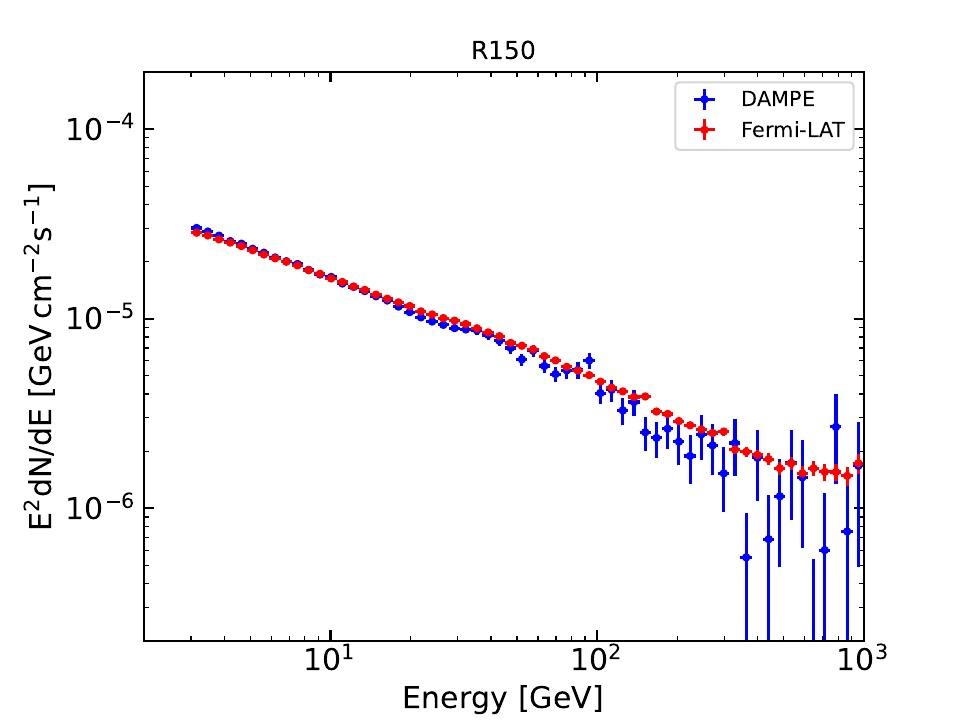}
\caption{The spectral energy distribution of the 4 ROIs based on 13-year Fermi-LAT and 5-year DAMPE observations.}
\label{fig:4sp}
\end{figure*}

Here we show the spectra of the 4 ROIs chosen to perform the line search. 
It can be seen that for the Fermi-LAT data, except for a few energies where there is spectral structure (which may be due to systematic errors or real signals), for most energies the spectrum is smooth and a power law can be used to fit a small segment of it. Although the DAMPE data have a much larger statistical fluctuation especially in the high energy end, its flux consistent with that by Fermi-LAT well. Note that the binned spectra in Fig.~\ref{fig:4sp} is just for visualization purpose, and the likelihood results in the main text are based on an unbinned analysis.

\section{Effective background and the likelihood with systematic term}
\label{ap:beff}
Systematic uncertainties may induce a false line signal or mask a true one in the fittings (\cite{Ackermann2013,Albert2014,Ackermann2015,Alemanno2021}). To account for such uncertainties, we adopt the same methodology as in \cite{Ackermann2015} to rewrite the likelihood equations. The systematic uncertainty term is added into Eq.~(1) and Eq.~(2) through
\begin{equation}
    \mathcal{L}\left(n=n_{\mathrm{sig}}\right)\rightarrow\ \mathcal{L}\left(n=n_{\mathrm{sig}}+n_{\mathrm{sys}}\right)\times\frac{1}{\sqrt{2\pi}\sigma_{\mathrm{sys}}}\exp{\left(-\frac{n_{\mathrm{sys}}^2}{2\sigma_{\mathrm{sys}}^2}\right)}.
\end{equation}
where $n_{\rm sig}$ is the true number of signal events and $n_{\rm sys}$ is the offset from the true value due to systematic effects.
The complete form of the (log) likelihood when considering systematic uncertainties is thus
\beq
\ln{\mathcal{L}_{\mathrm{tot}}\left(N_b,\Gamma,N_s,n_{\mathrm{sys}}\right)}=\ln{\mathcal{L}\left(N_b,\Gamma,N_s+n_{\mathrm{sys}}/\mathcal{E}\right)}+\ln{\frac{1}{\sqrt{2\pi}\sigma_{\mathrm{sys}}}}+\frac{n_{\mathrm{sys}}^2}{2\sigma_{\mathrm{sys}}^2}.
\label{eq:likefull}
\eeq
where $\mathcal{E}$ is the exposure and $\sigma_{\mathrm{sys}}=\delta f_{\rm sys}\times b_{\rm eff}$ is the width of the $n_{\rm sys}$ distribution in the fittings of control regions.

\begin{figure*}[h]
\centering
\includegraphics[width=0.5\textwidth]{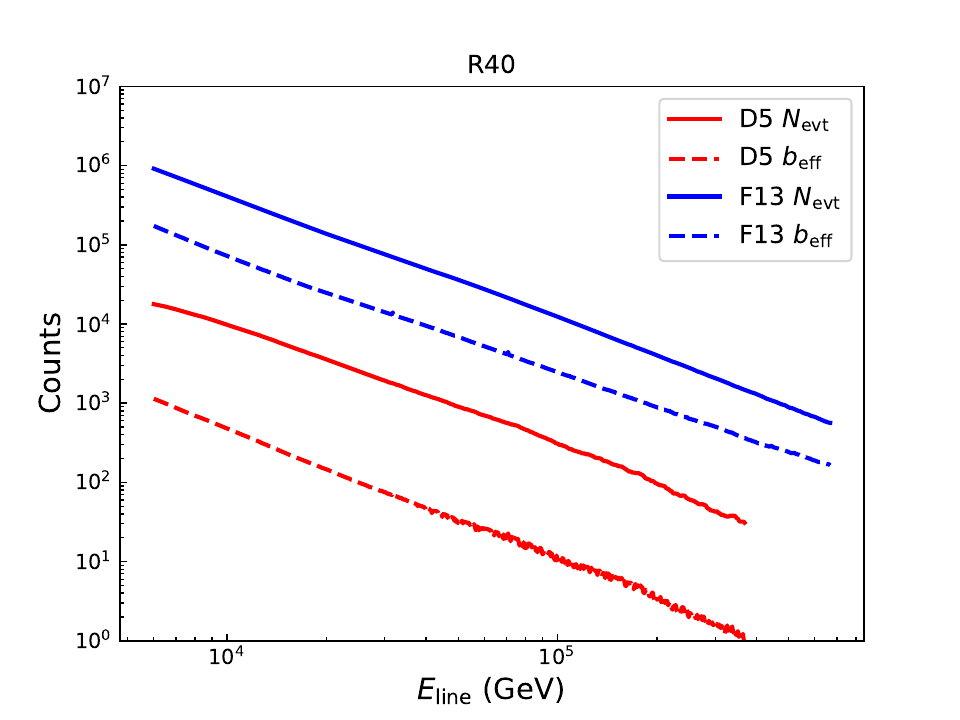}
\caption{The effective background $b_{\rm eff}$ and total counts in the considered window $n_{\rm evt}$ as a function the line energy $E_{\rm line}$ for Fermi-LAT and DAMPE data of R40 ROI.}
\label{fig:beff}
\end{figure*}

The concept of effective background $b_{\rm eff}$ is intended to represent the background that is most related to the signal search. It is not the total background in a whole energy window, but rather the part of the background beneath the signal. In the previous works of searching for spectral lines \cite{Ackermann2015,Alemanno2021}, the $b_{\rm eff}$ is defined as the number of background photons weighted by the counts distribution function of the signal (i.e., weight the background by how much it overlaps with the signal) and is expressed as
\begin{equation}
    b_{\mathrm{eff}}=\frac{n_{\rm evt}}{\left(\sum_i \frac{F_{\mathrm{sig}, i}^2}{F_{\mathrm{bkg}, i}}\right)-1}
\label{eq:beff}
\end{equation}
where $n_{\rm evt}$ is the total counts in the fit, $F_{\mathrm{sig}, i}$ and $F_{\mathrm{bkg}, i}$ are the binned normalized counts distribution functions for signal and background models within the $i$th energy bin. The summation in the denominator runs over 63 energy bins in each fit window.
When combining the DAMPE and Fermi-LAT data, Eq.~(\ref{eq:beff}) becomes
\begin{equation}
    b_{\text {eff}}=\frac{\sum_k n_k}{\left[\sum_k \sum_i \frac{n_k}{\left(n_f+n_d\right)} \frac{F_{\mathrm{sig}, i k}^2}{F_{\mathrm{bkg}, i k}}\right]-1}
\end{equation}
with $k=d$ or $f$ indicating the DAMPE or Fermi data set.
Fig.~\ref{fig:beff} shows the $b_{\rm eff}$ and $n_{\rm evt}$ in the window $\left[0.5E_{\rm line},1.5E_{\rm line}\right]$ as a function $E_{\rm line}$.

\section{Systematic uncertainties evaluation with control regions}
\label{ap:cr}
Systematic uncertainties in the line signal search primarily arise from the inaccurate modelling of the exposure and the approximation of the background spectrum as a power law \cite{Ackermann2013,Ackermann2015,Alemanno2021}.
In practice, one can estimate the level of systematic uncertainty through a data-driven method by fitting control regions dominated by astrophysical processes where no true signal is expected.
To do this, we define the fractional size (or called fractional signal), $f=n_{\mathrm{sig}}/b_{\mathrm{eff}}$.
In completely signal-free regions (control regions), the best-fit number of ``signal photons" obtained in the fit comes from two components: $f=\delta f_{\mathrm{stat}}+\delta f_{\mathrm{syst}}$, where the former and the latter are caused by statistical fluctuations and systematic effects, respectively.

In general, the systematic uncertainty is proportional to the observed photon counts and we assume that $f_{\mathrm{syst}}$ is energy invariant, so that $f_{\mathrm{syst}}$ is a constant for the $n_{\mathrm{sig,syst }}=f_{\mathrm{syst}}\times b_{\rm eff}$.
For the Fermi-LAT data, since the statistical uncertainty in the low-energy range is small/negligible, the systematic uncertainty dominates the total uncertainty.
The distribution of $f$ values in the fittings of control regions in the low-energy range can thus be used to determine the systematic uncertainties.
In the previous works, $\delta f_{\rm sys} \lesssim 1.5\%$ or $2.0\%$ were determined for the Fermi-LAT and the DAMPE, respectively \cite{Ackermann2015,Alemanno2021}.

\begin{figure*}[h]
\centering
\includegraphics[width=0.5\textwidth]{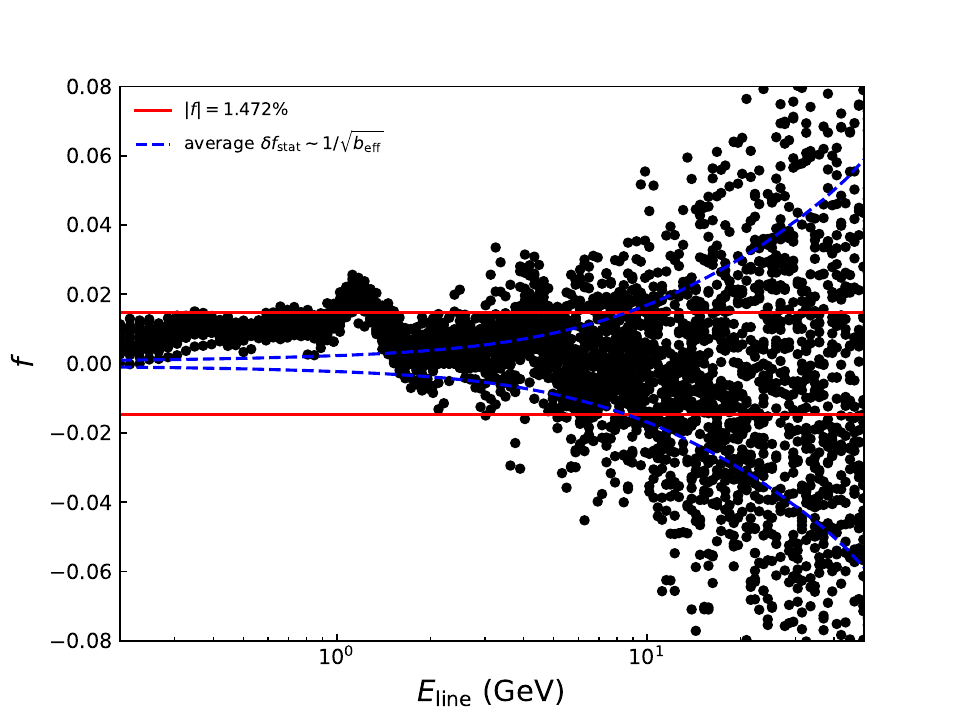}
\caption{The fractional signal $f$ in the control regions. Black points represent the observed $f$ values in 29 control regions along the Galactic plane. The blue dashed lines show the average standard deviation of the statistical term $\delta f_{\rm stat}$. From the control region fits, We obtain the level of the systematic uncertainties of $\delta f_{\rm sys}=1.472\%$ (horizontal lines).} 
\label{fig:cr}
\end{figure*}

Considering that the Fermi-LAT data used in this work have been updated compared to the previous analysis (we use P8R3 data while \cite{Ackermann2015} used the first P8 version), we refit the control regions to check if the previous systematic errors derived for Fermi-LAT data still applied. In fitting of the control regions, we adopt the same sliding window method as used in the line searches, with the same line energies determined by the instrument energy resolution. Then the energy windows are defined with a width of $[0.5 E_{\rm line}, 1.5 E_{\rm line}]$. The Galactic plane is chosen as the control region.
We scan for spectral lines in 29 box-shape control regions ($10^\circ\times10^\circ$ each) along the Galactic plane ($|b|<5^\circ$). These regions are selected starts from $l=35^\circ$ and ends at $l=325^\circ$ with a step width of $10^\circ$.
We calculate the $f$ values that contain 68\% of the control region fits, $\delta f_{68}(E)$, in a small energy range ($\pm$10\%). To be conservative, the largest $\delta f_{68}$ value observed in the energy scan (100 MeV$<E_\gamma<$5 GeV) is chosen as the estimate for the systematic uncertainty.
The results of our fits of control regions are shown in Fig.~\ref{fig:cr} and we find that the obtained $\delta f_{\rm sys}$ value are $\sim1.47\%$ which does not differ with Ref.~\cite{Ackermann2015}.
The width of the Gaussian term in Eq.~(4) and (\ref{eq:likefull}) is obtained by $\sigma_{\rm sys}=\delta f_{68}\times b_{\rm eff}$.
Note that the derived $f$ is biased to positive value at $E_{\rm line} \lesssim$ 5 GeV, as shown in Fig.~\ref{fig:cr}. This could relate to the uncertainty induced by the power-law background assumption. An alternative background model, for example a log-parabola, may lead to stronger constraints on dark matter parameters.

In addition, there are other systematic uncertainties that cannot be considered and corrected by this method. For example, the uncertainty caused by cosmic-ray residual contamination, and the systematic bias of the overall calibration of the effective area \cite{Ackermann2013}. Because the CR residual background is not dominant in the Galactic plane region, it cannot be estimated by the analysis of the control regions in the Galactic plane. It has been shown that \cite{Ackermann2013,Ackermann2015}, this systematic uncertainty only has an impact on the analysis of large sky regions (e.g., R90 and R150 ROIs) and the impact is sub-dominant.
For the overall uncertainty of the effective area, the Fermi-LAT Collaboration has estimated that it is $\sim5\%$ for the energies $>$100 MeV and $<$100 GeV \footnote{\url{https://fermi.gsfc.nasa.gov/ssc/data/analysis/scitools/Aeff_Systematics.html}, see also Ref.~\cite{Fermi-LAT:2012fsm}.}.
This uncertainty generally does not cause pseudo-signal or weaken a signal, but makes the derived fluxes (and thus the constraints on $\left<\sigma v\right>$ and $\tau$) have an uncertainty of 5-10\%.

\section{Expected sensitivity of spectral line search}
\label{ap:expsens}
The definition of the $b_{\rm eff}$ essentially represents the uncertainty of the signal counts in an ideal case (where parameters are completely independent and the off-diagonal elements of the covariance matrix are zero) when performing the signal search.
So we can use $b_{\rm eff}$ to predict the sensitivity of spectral line search. The expected sensitivity considering only the statistical uncertainty can be expressed as
\begin{equation}
    S=\sqrt{b_{\rm eff}}/\mathcal{E},
    \label{eq:exps1}
\end{equation}
while with the systematic uncertainty included it is
\begin{equation}
    S=(\sqrt{b_{\rm eff}}+\delta f_{\rm sys}b_{\rm eff})/\mathcal{E},
    \label{eq:exps2}
\end{equation}
where $\mathcal{E}$ is the exposure.

\begin{figure*}[h]
\centering
\includegraphics[width=0.4\textwidth]{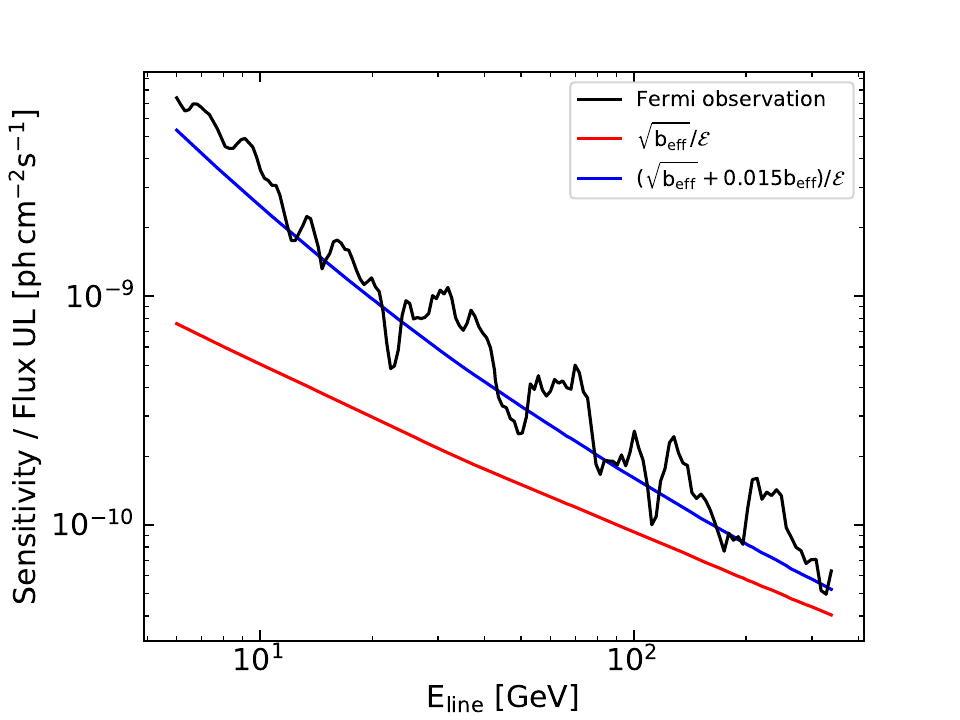}
\includegraphics[width=0.4\textwidth]{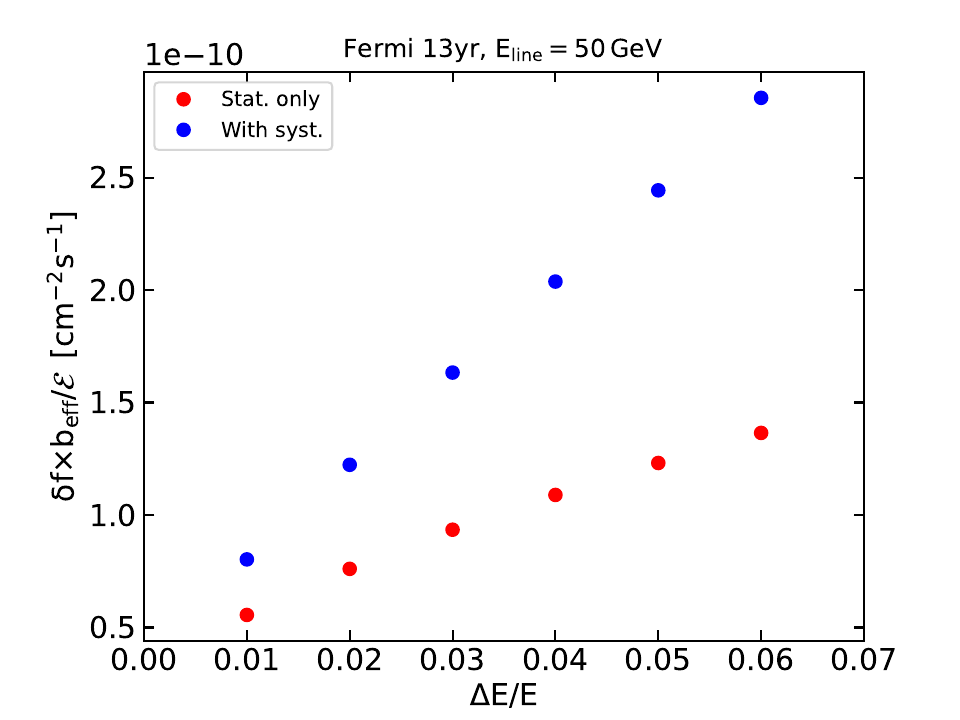}\\
\includegraphics[width=0.4\textwidth]{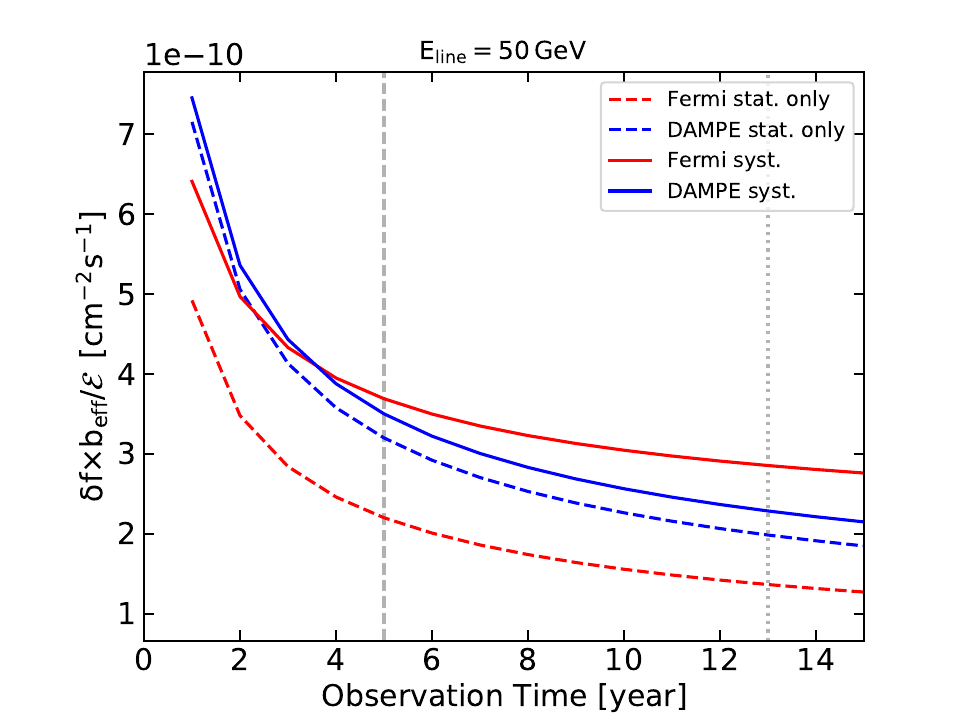}
\includegraphics[width=0.4\textwidth]{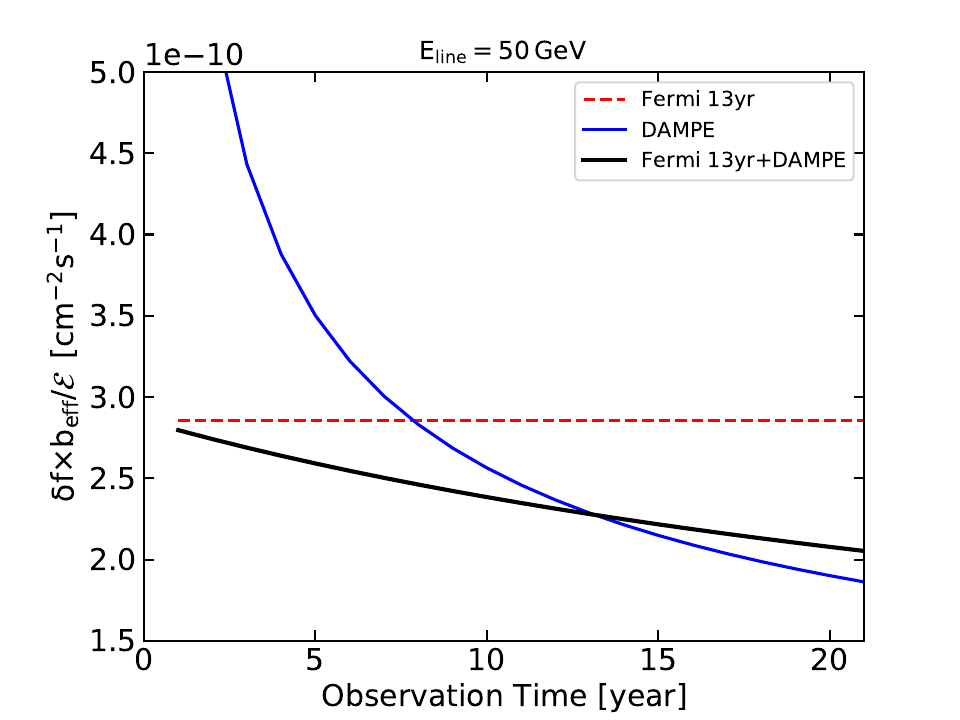}
\caption{{\it Upper left panel:} Expected sensitivities at different line energies for 13 years of Fermi-LAT observations of R40 ROI (blue and red lines) and a comparison with the flux upper limits derived from real observations (black line). {\it Upper right panel:} This plot shows how the sensitivity relies on the energy resolution $\Delta E/E$ of the instrument in the case of considering/not considering systematic uncertainties. {\it Lower left panel:} The change in the line search sensitivity as a function of observation time. The two vertical lines denote the observation time of the DAMPE and Fermi-LAT data adopted in our work. {\it Lower right panel:} Predicted sensitivity for the analysis combining 13 years of Fermi-LAT data and different years of DAMPE data (black line). The blue and red lines are for the sensitivities based on the Fermi-LAT or DAMPE data alone.}
\label{fig:sens}
\end{figure*}

Using Eqs.~(\ref{eq:exps1}) and (\ref{eq:exps2}) we derive the expected sensitivity as a function of the line energy $E_{\rm line}$, energy resolution $\Delta E/E$, and photon statistics. Fig.~\ref{fig:sens} shows the results of the expected sensitivity for the spectral line search of the R40 ROI.
If not specified in the figures, our baseline parameters adopt the Fermi-LAT 13-year exposure of R40 at $50\,{\rm GeV}$ ($\sim5.3 \times 10^{11}\,{\rm cm^{2}\,s}$) and the total photon count $n_{\rm evt}$ for the window of $[25\,{\rm GeV},75\,{\rm GeV}]$.
For DAMPE we consider the exposure of 5 years of observation ($\sim1.5 \times 10^{10}\,{\rm cm^{2}\,s}$). Fermi-LAT and DAMPE energy resolutions are assumed to be 6\% and 1\% (68\% containment, both assuming Gaussian energy dispersion for simplicity), respectively.
We consider a power-law background model with an index of $-2.5$ and a systematic uncertainty of $\delta f_{\rm sys} = 1.5\%$ is introduced.

The upper-left panel shows the expected sensitivities at different energies for 13 years of Fermi-LAT observations and compares them with the results based on real observations. It can be seen that the predictions using $b_{\rm eff}$ are in good agreement with the flux upper limits obtained from the real observations, indicating that the equation we used to predict the sensitivity is feasible, and also justify our search results.
The upper-right panel and the bottom panels show how the sensitivity changes as the energy resolution improves or as the amount of data increases. As expected, both increasing data amount (depending on the instrument effective area and total observation time) and improving energy resolution could lead to stronger constraints / better detection sensitivity of gamma-ray lines. For the energy resolution, there is a $\sim 3$ times improvement of the line search sensitivity from energy resolution $6\%$ to $1\%$.

For the relation between sensitivity and data accumulation, if systematic uncertainties are ignored, from Eq.~(\ref{eq:exps1}) it is easy to see that $b_{\rm eff}\propto n$, so that $S\propto1/\sqrt{\mathcal{E}}$.
In such a case, the Fermi-LAT generally has better sensitivity than the DAMPE, due to its larger effective area and longer exposure time.
However, when including the systematic uncertainty term in the likelihood, as the data accumulate the statistical error becomes smaller and $\sqrt{b_{\rm eff}}<\delta f_{\rm sys}b_{\rm eff}$, so the sensitivity will be dominated by the systematic term, i.e. $S\approx \delta f_{\rm sys}b_{\rm eff}/\mathcal{E}\propto\delta f_{\rm sys}=const$. Therefore, for the right part of the red solid line of Fig.~\ref{fig:sens}, the sensitivity will not change significantly with the observation time. The sensitivity of DAMPE will instead be stronger than the Fermi-LAT, for its higher energy resolution which makes the $b_{\rm eff}$ smaller.
Finally for the sensitivity of the joint analysis, it can be seen in the lower right panel, which shows combining the DAMPE data does improve the results compared to the Fermi-LAT only ones. At $E_{\rm line}=50\,{\rm GeV}$, adding 5 years of DAMPE data will improve the sensitivity by about 10\%, demonstrating the necessity of the joint analysis for improving the sensitivity.

From the results here, we can also estimate the detection sensitivity of future telescopes with larger effective areas. For example, the Very Large Area Gamma-ray Space Telescope (VLAST; \cite{2022AcASn..63...27F}) is planned to have an effective area 5 times that of Fermi-LAT and an energy resolution comparable to the DAMPE (up to 1\% at 10~GeV). According to the simulation results, its observations will improve the current most stringent limits by a factor of $\sim5$.

\section{Results based on EDISP3 data}
\label{ap:ed3}
The Fermi-LAT data allows for the subdivision of events into different event types \footnote{\url{https://fermi.gsfc.nasa.gov/ssc/data/analysis/documentation/Cicerone/Cicerone_Data/LAT_DP.html}
}. The EDISP (energy dispersion) event types partition the data according to the quality of the energy reconstruction. There are four EDISP event types, ranging from EDISP0 to EDISP3, where EDISP3 is the subgroup of events in Fermi-LAT data that have the best energy reconstruction quality, i.e. the lowest uncertainty in energy measurement.
All EDISP event types have a similar number of events in each logarithmic energy bin and are mutually exclusive.

The following figures show the results from the analysis using the EDISP3 data. For the Fermi-LAT only analysis, in the low energy regime because the statistics are sufficient and the systematic errors become important, increasing the energy resolution (which reduces the $b_{\rm eff}$ and $\sigma_{\rm sys}$) does improve the results of the analysis. In the high energy end, on the other hand, the results are a bit better for the Front+Back data (especially the R150 case) due to the larger statistics compared to EDISP3. In the joint analysis, the results of EDISP3 and Front+Back show no significant difference in the low-energy range while at high energies the Front+Back results are mildly better.

\begin{figure*}[h]
\centering
\includegraphics[width=0.8\textwidth]{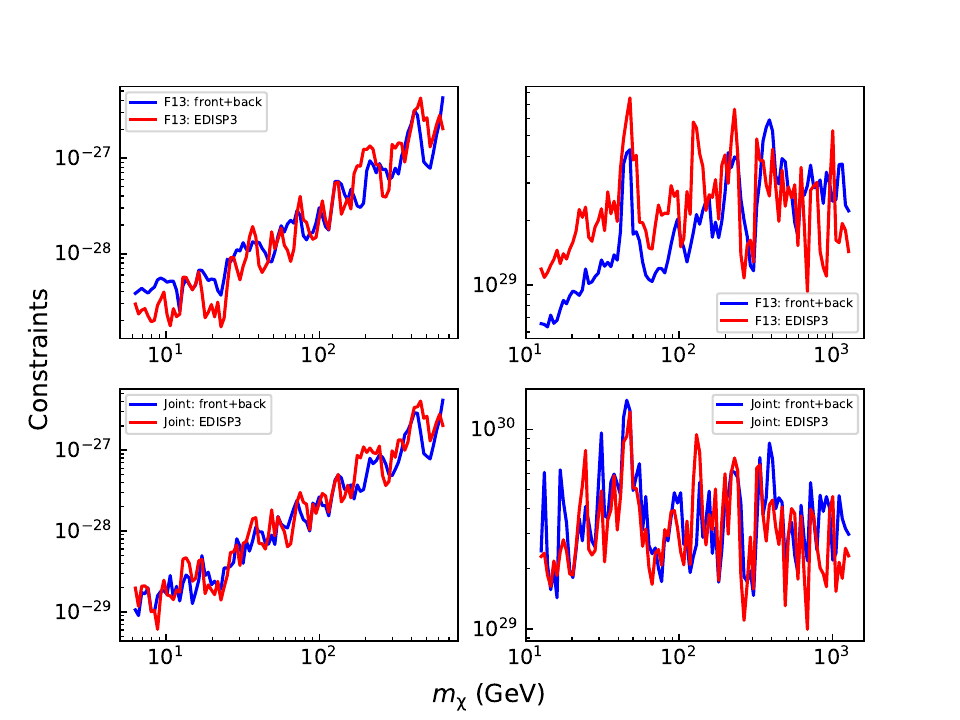}
\caption{Comparisons of constraints derived using the EDISP3 and Front+Back data of Fermi-LAT. The upper (lower) panels are for the results based on Fermi-LAT data alone (from the joint analysis).}
\label{fig:ed3}
\end{figure*}

\section{Results ignoring systematic uncertainties}
\label{ap:nosys}

Systematic uncertainties may affect the results of the line signal search. To quantitatively incorporate them into the analysis, the fractional signal $f_{\rm sys} \equiv n_{\rm sig} / b_{\rm eff}$ is used as elaborated in Appendix \ref{ap:beff} and \ref{ap:cr}, in which $n_{\rm sig}$ and $b_{\rm eff}$ are the photon counts of the signal and effective background, respectively \citep{Ackermann2013,Albert2014,Ackermann2015}. For the PASS 8 data of the Fermi-LAT, the total systematic uncertainty given by analyzing the control regions is estimated to be $\delta f_{\rm sys} \lesssim 1.6\%$ \citep{Ackermann2015}. Similarly, for the DAMPE data, the overall systematic uncertainty is estimated to be $\delta f_{\rm sys} \lesssim 2.0\%$ \citep{Alemanno2021}.

\begin{figure*}
\centering
\includegraphics[width=0.4\textwidth]{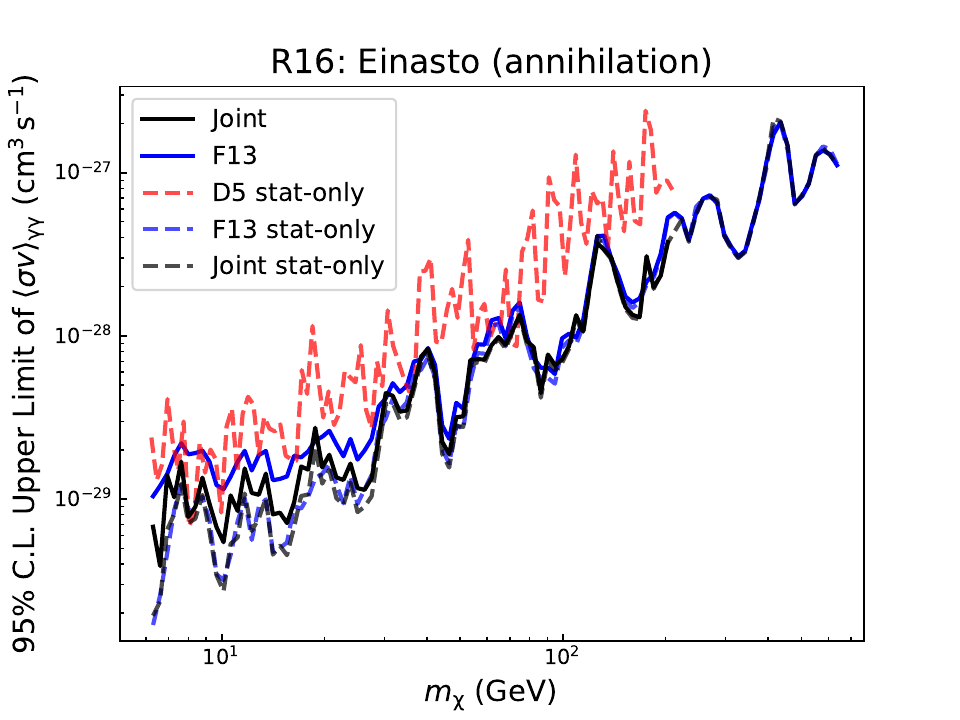}
\includegraphics[width=0.4\textwidth]{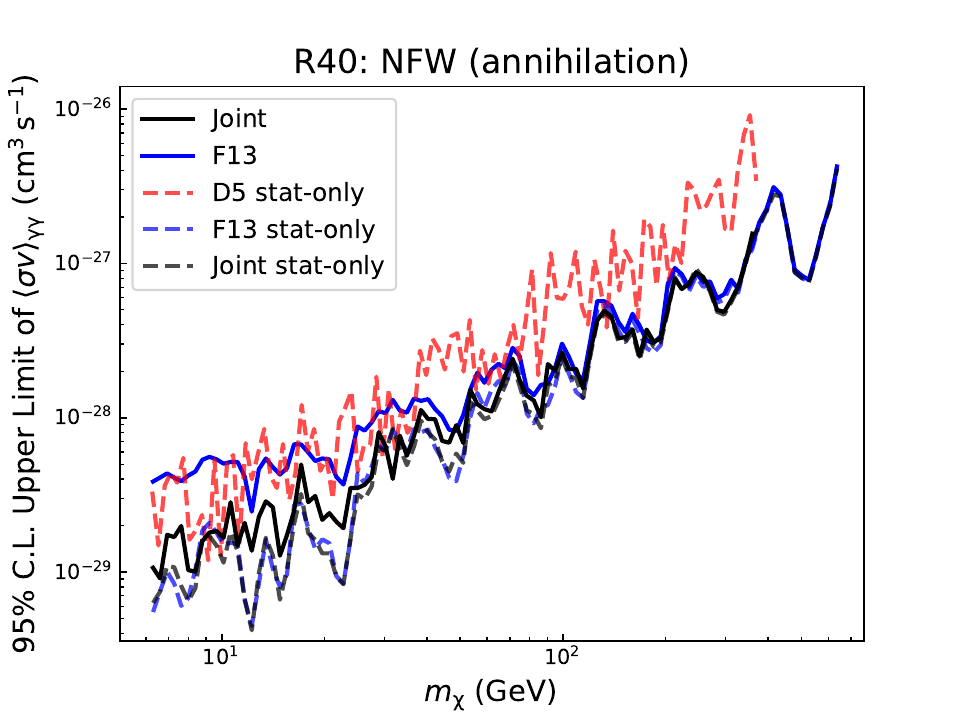} \\
\includegraphics[width=0.4\textwidth]{fig3c.pdf}
\includegraphics[width=0.4\textwidth]{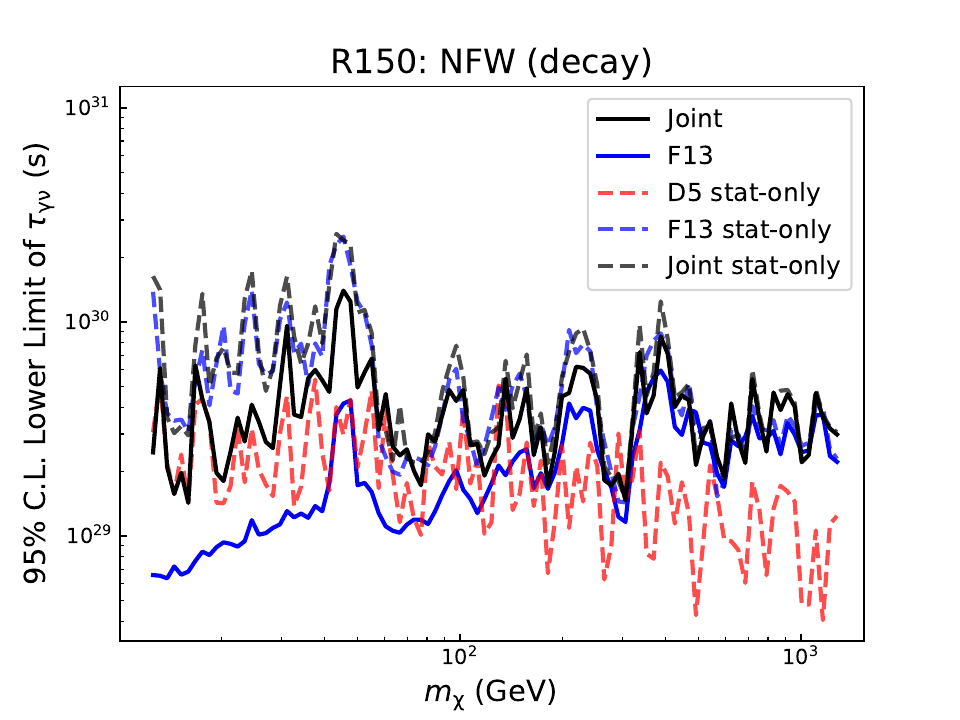} \\
\caption{Constraints on $\langle{\sigma v}\rangle_{\gamma\gamma}$ and $\tau_{\gamma \nu}$ from the analysis ignoring systematic uncertainties and comparisons to our nominal results which take systematic effects into account. The {\it Joint stat-only} results and the {\it F13 stat-only} results are only slightly different.}
\label{fig:stat2}
\end{figure*}

In our nominal results (Fig.~2 of the main text), we have used the same approach as in \cite{Ackermann2015} to account for the systematic uncertainties. Here we further present the results that ignore the systematic uncertainties in the likelihood analysis, which can help us to understand the role of the data from the two instruments in the joint analysis. The results are shown in Fig.~\ref{fig:stat2}.
As is shown, constraints derived from the 13 years' Fermi-LAT data are overall much stronger than the D5 results, benefited from the larger effective area and longer operation time of the Fermi-LAT. Consequently, the constraints from the joint analysis of the Fermi-LAT and DAMPE data are dominated by the Fermi-LAT data and are only slightly stronger than the single F13 results.

Comparing the results of considering systematic uncertainty with the {\tt stat-only} ones here, it can be seen that including the systematic uncertainty prevents the Fermi-LAT results from improving with the data accumulation, especially in the low-energy range. In contrast, although the DAMPE data are statistically subdominant, including DAMPE data in the joint analysis can effectively reduce the sensitivity loss due to the consideration of systematic uncertainty because of its higher energy resolution (and therefore smaller $b_{\rm eff}$).

\section{Data availability}
The Fermi-LAT data and software underlying this article are publicly available in the Fermi Science Support Center, at \url{https://fermi.gsfc.nasa.gov/ssc/data/access/} (data) and \url{https://github.com/fermi-lat/Fermitools-conda} (software). The DAMPE data and software are available at \url{https://dampe.nssdc.ac.cn/dampe/dataquerysc.php} (data) and \url{https://dampe.nssdc.ac.cn/dampe/dampetools.php} (software).

\bibliographystyle{apsrev4-1-lyf}
\bibliography{references.bib}

\end{document}